\documentclass[aps,prl,twocolumn,superscriptaddress,address,amsmath,amssymb,showpacs]{revtex4}

\usepackage{graphicx}
\usepackage{subfigure}  
\usepackage{hyperref}

\newcommand{\etal}{{\it et al.}\/}
\newcommand{\gtwid}{\mathrel{\raise.3ex\hbox{$>$\kern-.75em\lower1ex\hbox{$\sim$}}}}
\newcommand{\ltwid}{\mathrel{\raise.3ex\hbox{$<$\kern-.75em\lower1ex\hbox{$\sim$}}}}
\begin{document}


\title{Changes in the self-energy and $d$-wave pairing strength with doping in overdoped La$_{2-x}$Sr$_x$CuO$_4$}

\author{Thomas Dahm}
\affiliation{Universit\"at Bielefeld, Fakult\"at f\"ur Physik, 
Postfach 100131, D-33501 Bielefeld, Germany}

\author{D.J.~Scalapino}
\affiliation{University of California, Physics Department, Santa Barbara, CA 93106-9530, USA}

\date{\today}

\begin{abstract}
Angle resolved photoemission spectroscopy (ARPES) studies of the overdoped
cuprate superconductor La$_{2-x}$Sr$_x$CuO$_4$ find only small changes in the near nodal electron
self energy over a spectral range of several hundred meV as the doping increases
from $x=0.2$ to $x=0.3$ and the superconducting transition temperature
$T_c$ decreases from 32K to 0K.  These measurements
put constraints on the structure of the electron-electron interaction. Here we
show that a spin-fluctuation interaction leads to behavior which is
consistent with these experimental results.
\end{abstract}

\pacs{
74.72.-h, 	
74.25.Jb, 	
79.60.-i, 	
74.20.Mn 	
}

\maketitle


A possible way to identify the pairing interaction responsible for
superconductivity in a given material is to investigate the
structure in the effective self-energy extracted from 
angle resolved photoemission spectroscopy (ARPES) measurements.  
\cite{Valla, Lanzara, Kaminski, Johnson, Zhou, Zhou2, Borisenko, Meevasana, Kordyuk, Yoshida, Peng, Yun, Park}
Here the basic premise is that the interaction that determines the
normal self-energy must also play a central role in the
anomalous (pairing) self-energy. 
However, in the cuprates, the interaction is strongly anisotropic
leading to a $d$-wave pairing state.
Then, the normal self-energy involves a projection of the electron-electron 
interaction which is determined by the one-electron Green's function itself 
while the strength of the $d$-wave pairing interaction depends upon a 
$d$-wave $(\cos k_xa - \cos k_ya)$ projection of the singlet part of the 
electron-electron interaction. Thus, the information about the pairing 
interaction provided by ARPES in the normal state is indirect. \cite{Yun}
Nevertheless, as the superconducting transition
temperature in the cuprates strongly changes with doping, 
the doping dependence of the ARPES data provides additional 
information and can put different scenarios for the pairing 
interaction under test.
Recently Park \etal\ \cite{Park} have reported ARPES measurements on
overdoped La$_{2-x}$Sr$_x$CuO$_4$ (LSCO) with different doping levels
that raise questions regarding the origin
of the pairing interaction. In the present work we provide
theoretical calculations within a spin-fluctuation scenario
and study the doping dependence of the nodal electronic
structure and the strength of the $d$-wave pairing interaction.

In the work by Park \etal\ \cite{Park}, the momentum distribution curve
(MDC) dispersion for nodal
momentum cuts, as well as several nearby cuts, were measured for dopings of
$x=0.20$ ($T_c=32$~K) and $x=0.30$ ($T_c=0$~K). Then taking a linear bare
band dispersion connecting the MDC peak positions between $-0.2$ eV and 0.0 eV,
the real part of an effective self energy Re~$\Sigma_{\rm eff}(k,\omega)$ was
determined for each doping. The idea was to see whether the changes in
Re~$\Sigma_{\rm eff}(k,\omega)$ over this 200~meV spectral region correlated
with the disappearance of superconductivity. For both dopings a kink appeared
in the disperson of the MDC peak near 70~meV giving rise to a broad peak in
Re~$\Sigma_{\rm eff}(k,\omega)$ as a function of energy $\omega$. 
The amplitude of the peak in Re~$\Sigma_{\rm eff}(k,\omega)$
was observed to decrease by only about 30\% between the superconducting $x=0.2$
($T_c=32$K) and non-superconducting $x=0.3$ material. The authors noted that
the broad nature of the peak indicated coupling to a spectrum of modes but
concluded that the observed change in Re~$\Sigma_{\rm eff}(k,\omega)$ between
the $x=0.2$ and 0.3 samples was insufficient to account for the decrease in $T_c$
if the pairing interaction arose from modes in this spectral range.

Here we re-examine this conclusion based on a fluctuation exchange
(FLEX) \cite{Bickers,DT} calculation of the single particle spectral function
for a Hubbard model of overdoped LSCO. Although FLEX is inherently a weak coupling
approximation, we believe that it can provide a useful approximation for the
overdoped regime.
Since we are interested in determining the spectral function and self energy
as functions of frequency, we will work on the real frequency axis as
described in Ref.~\onlinecite{DT}. This avoids the need of analytic continuation
from imaginary Matsubara frequencies to the real axis, which can be numerically
unstable.


Within FLEX the imaginary part of the self-energy $\Sigma$
is obtained from
\begin{eqnarray}
 \mbox{Im}\;\Sigma\left(k,\omega\right)&=&
 -\frac{1}{N} \sum_q \int_{-\infty}^{\infty}d\Omega\,
    \left[n\left(\Omega \right)+f\left(\Omega-
    \omega \right)\right]\nonumber\\
    &&\times \mbox{Im}\;\Gamma\left(q,\Omega\right) 
    A\left(k-q,\omega-\Omega \right).
    \label{Eqflex1} 
\end{eqnarray}
Here, $f$ and $n$ are the usual Fermi and Bose functions, 
respectively, and $N$ is the number of lattice sites. The vertex function $\Gamma$
includes the interactions due to spin and charge
fluctuations and is given by
\begin{eqnarray}
  \Gamma & = & \frac{3}{2} \ 
  \frac{U^2 \chi_{0}}{1-U\chi_{0}} + \frac{1}{2}\  \frac{U^2 
  \chi_{0}}{1+U\chi_{0}} -  U^2 \chi_{0}  \label{Eqflex2} 
\end{eqnarray}
where the last term removes a double counting. $\chi_{0}$ is calculated 
selfconsistently from
\begin{eqnarray}
 \mbox{Im} \; \chi_{0} \left( q, \Omega \right) &=& 
    \frac{\pi}{N} \sum_k \int_{-\infty}^{\infty} d\omega \,
    \left[ f \left( \omega \right) - f \left( \omega + 
    \Omega \right) \right] \nonumber \\
  && \times A \left( k ,\omega \right) 
    A \left( k+q,\omega + \Omega \right)  \label{Eqflex3} 
\end{eqnarray} 
with $A(k,\omega)$ the single particle spectral weight in the normal state given by
\begin{eqnarray}
  A \left( k ,\omega \right) & = & - \frac{1}{\pi} \mbox{ Im } 
  \frac{1}{\omega - \epsilon_k - \Sigma \left( k, \omega \right)} 
  \label{Eqflex4} 
\end{eqnarray}
The real parts of Eqs.~(\ref{Eqflex1}) and (\ref{Eqflex3})
are calculated from Kramers-Kronig transformations. In the FLEX approximation,
Equations (\ref{Eqflex1}) to (\ref{Eqflex4}) comprise a set
of coupled integral equations that are iterated until
a selfconsistent solution is obtained.

\begin{figure}[t]
\includegraphics[width=0.6 \columnwidth]{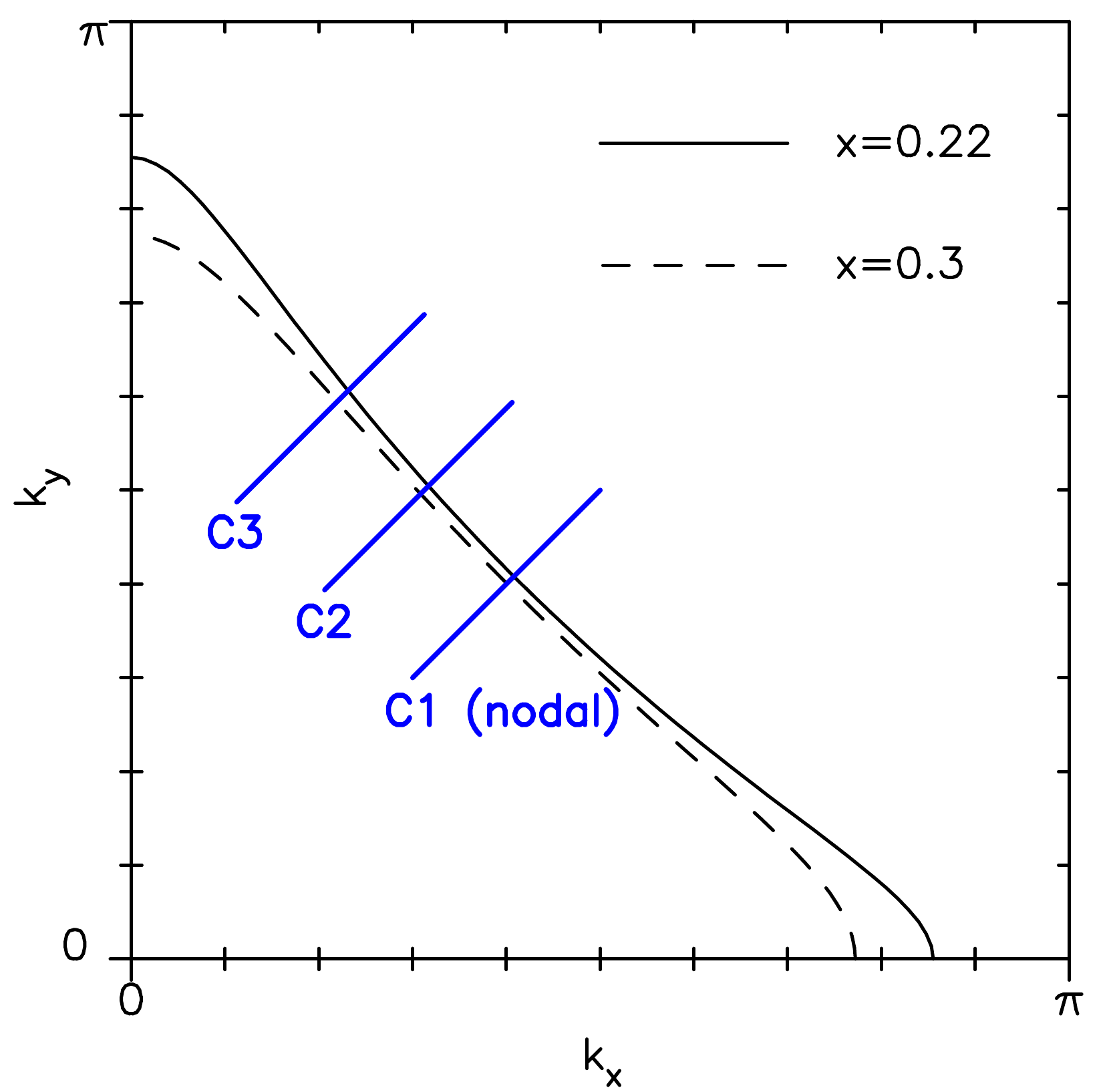}
\caption{(Color online) The Fermi surfaces for $x=0.22$ (solid) and 0.30 (dashed) obtained
from tight-binding parameters given by Yoshida \etal\ \cite{Yoshida}
Here C1 is a nodal momentum cut and the C2 and C3 cuts are offset from the
nodal cut by 0.1875 $\pi/a$ in the $k_y$ direction.\label{fig:1}}
\end{figure}

For overdoped La$_{2-x}$Sr$_x$CuO$_4$, information on the renormalized bandstructure
has been obtained from tight-binding fits to ARPES data by Yoshida \etal\ \cite{Yoshida} 
Their tight-binding fit is of the form
\begin{eqnarray} 
\bar{\epsilon}_k/t &=& -2 \left[ \cos \left( k_x a \right) +
\cos \left( k_y a \right) \right] - 4 t' \cos \left( k_x a \right)
\cos \left( k_y a \right) \nonumber \\ & &  - 2 t'' \left[ \cos \left( 2 k_x a \right) + 
\cos \left( 2 k_y a \right) \right] - \mu \label{eq:tbfit}
\end{eqnarray}
and for $x=0.22$ and 0.3 gives the Fermi surfaces shown in Fig.~\ref{fig:1}.
In the FLEX calculation, we adjust the bandstructure $\epsilon_k$ in such a way
that the renormalized Fermi surface at each doping is
fixed to the ARPES determined Fermi surface for that
doping. This is done by requiring that at each
iteration of the FLEX calculation the quantity
$\epsilon_k + \mbox{Re} \; \Sigma \left( k, \omega=0 \right)$
remains equal to $\bar{\epsilon}_k$ and amounts to setting \cite{Ikeda,suppinf}
\begin{eqnarray}
  \lefteqn{A \left( k ,\omega \right) =} \nonumber \\
 & & - \frac{1}{\pi} \mbox{ Im } 
  \frac{1}{\omega - \bar{\epsilon}_k - \Sigma \left( k, \omega \right)
  + \mbox{Re} \; \Sigma \left( k, \omega=0 \right)} 
  \label{Eqflex4p} 
\end{eqnarray}
In this way we ensure that the selfconsistent Fermi surface of the FLEX
calculation is identical to the experimental one. In the following we will
concentrate on the doping levels $x=0.22$ and $x=0.3$, because tight-binding
fits are available for these doping levels from Yoshida \etal\ For $x=0.22$ the fit
parameters are $t'=-0.13$, $t''=0.065$, and $\mu=-0.88$.
For $x=0.3$ the fit parameters are $t'=-0.12$, $t''=0.06$, and $\mu=-0.99$.\cite{Yoshida}
Our calculations are done in energy units of $t$. The parameter $t$
is then determined at the end of the calculation in such a way that the nodal
MDC peak position of our FLEX calculation agrees with the experimental 
one from Park \etal\  \cite{Park} at $\omega=-200$~meV. In this way our value
of $t$ represents an unrenormalized hopping in contrast to the
renormalized value of $\bar{t}=250$~meV found by Yoshida \etal\cite{Yoshida}
For our numerical calculations we have chosen a moderate value
of $U/t=3$ to stay in a weak coupling regime where FLEX can be assumed
to give reliable results. \cite{QMC} We will show results for the
self-energy and spectral functions along the same momentum cuts C1-C3 as
Park \etal\  \cite{Park}, shown in Fig.~\ref{fig:1}.

\begin{figure}[t]
\begin{center}
\includegraphics[width=0.9 \columnwidth]{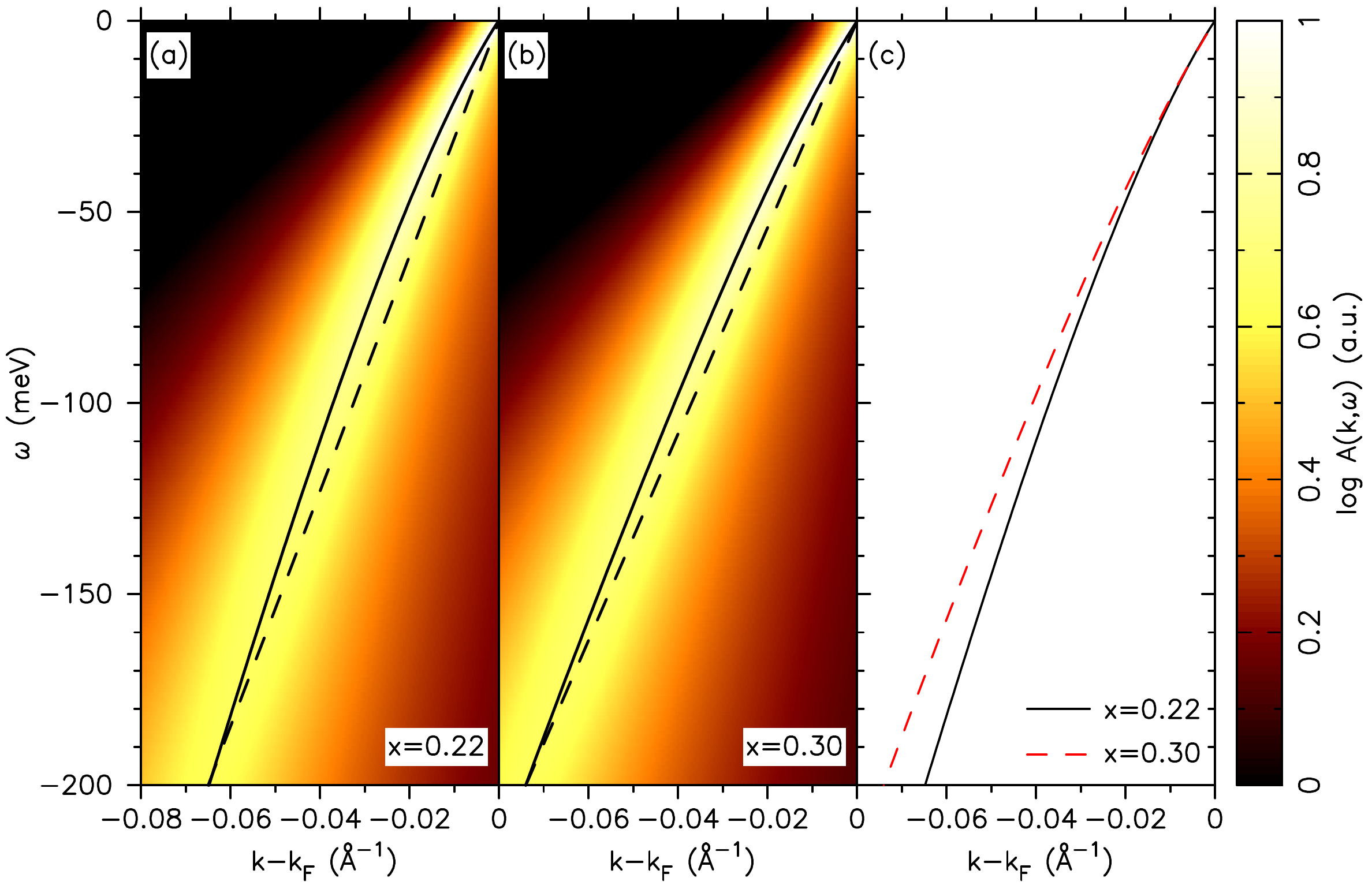}
\caption{(Color online) Single particle spectral function $A(k,\omega)$ for
  the nodal momentum cut C1
at dopings $x=0.22$ and $x=0.3$. In (a) and (b) the solid line shows the 
MDC peak
dispersion and the dashed line is a linear dispersion connecting the MDC peaks
at 0.0 eV and $-0.2$ eV. In (c) the MDC peak for $x=0.22$ (solid) is
compared with that for $x=0.3$ (dashed) and one sees that the dispersion is the
same at low energies. \label{fig:2}}
\end{center}
\end{figure}


{\it Results} - The single particle spectral weights obtained from the FLEX calculation for
$x=0.22$ and $x=0.3$ along the nodal C1 cut are shown in Figs.~\ref{fig:2}a
and b, respectively. Here the $y$-axis denotes energy in units of meV 
and the $x$-axis the
wavevector in units of inverse Angstroms. A lattice constant $a=3.79$\AA\ was
used to set the wavevector scale. The solid curves in Figs.~\ref{fig:2}a
and b show the MDC peak and the
dashed line is a linear dispersion connecting the peaks at $\omega=0.0$~eV and
$-0.2$~eV. For $x=0.22$, we find the bare $t=500$~meV while for $x=0.3$, 
$t=415$~meV. The dispersions of the MDC peak for the two dopings are compared
in Fig.~\ref{fig:2}c.

In Fig.~\ref{fig:2} one sees that there is a kink in the dispersion of the MDC
peak at approximately $-70$~meV for both $x=0.22$ and 0.3. 
The appearance of a nodal kink in the normal state of spin-fluctuation models 
has been discussed by several authors before. \cite{Manske,Chubukov,Dahm} 
For the FLEX approximation this feature was pointed out by Manske \etal\cite{Manske} The
renormalized nodal Fermi velocity obtained from the slope of the dispersion in
the low energy region is 1.94 eV\AA\ for $x=0.3$ and 1.98 eV\AA\ for $x=0.22$,
consistent with the behavior of the Fermi velocity reported by Zhou \etal \cite{Zhou}
and also seen in Ref.~\onlinecite{Park}. This is clearly seen
in Fig.~\ref{fig:2}c which shows the MDC peak dispersion for both dopings.
Following Park \etal\ \cite{Park}, we
define the real part of an effective self energy Re~$\Sigma_{\rm eff}(k,\omega)$
by the $\omega$-deviation of the MDC peak from the dashed line in
Fig.~\ref{fig:2}. Note, that this effective low energy quantity is different 
from the full FLEX self energy in Eq.~(\ref{Eqflex1}).
\begin{figure}[t]
\begin{center}
\subfigure{
\includegraphics[width=0.45 \columnwidth]{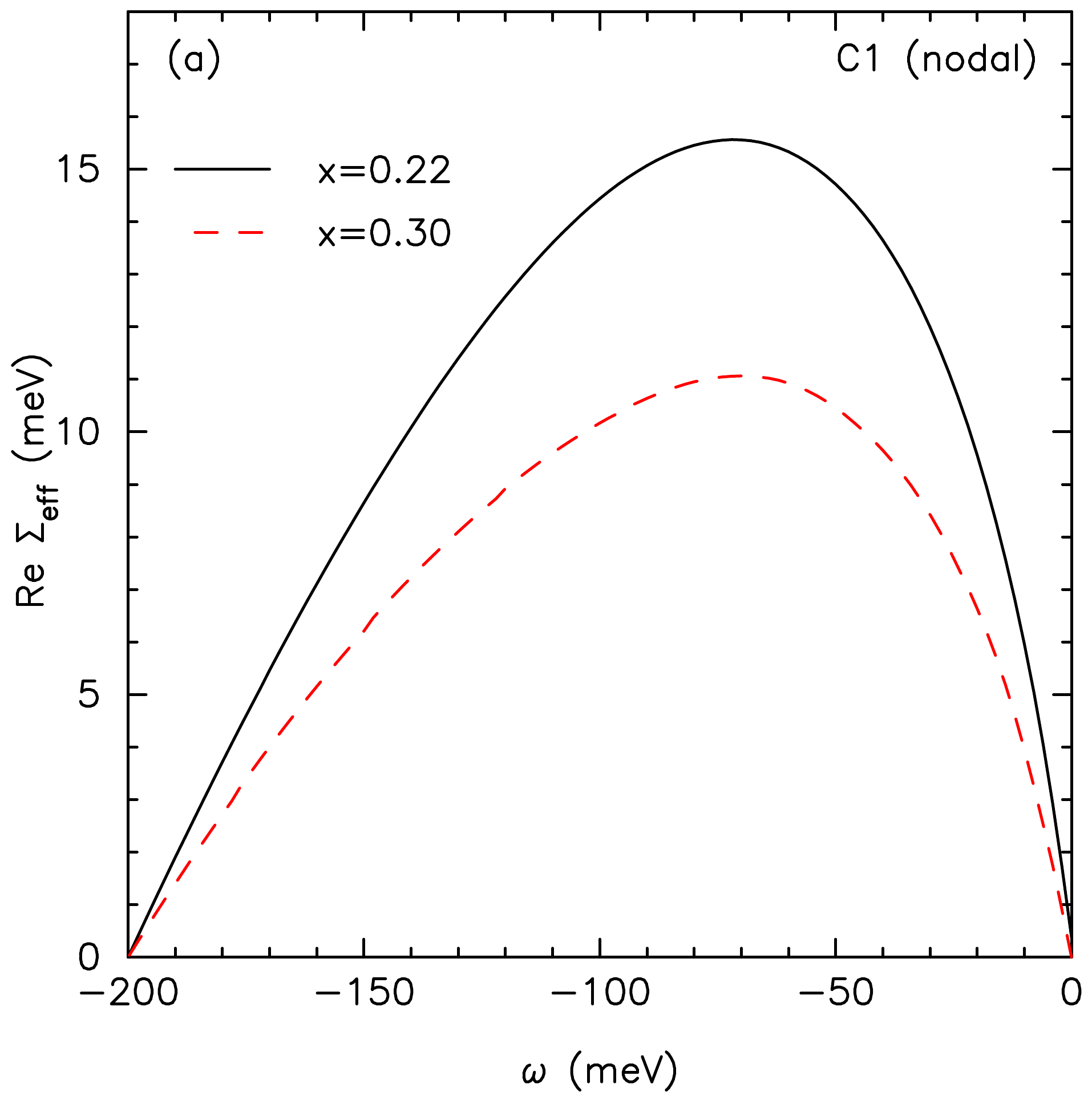}}
\subfigure{
\includegraphics[width=0.45 \columnwidth]{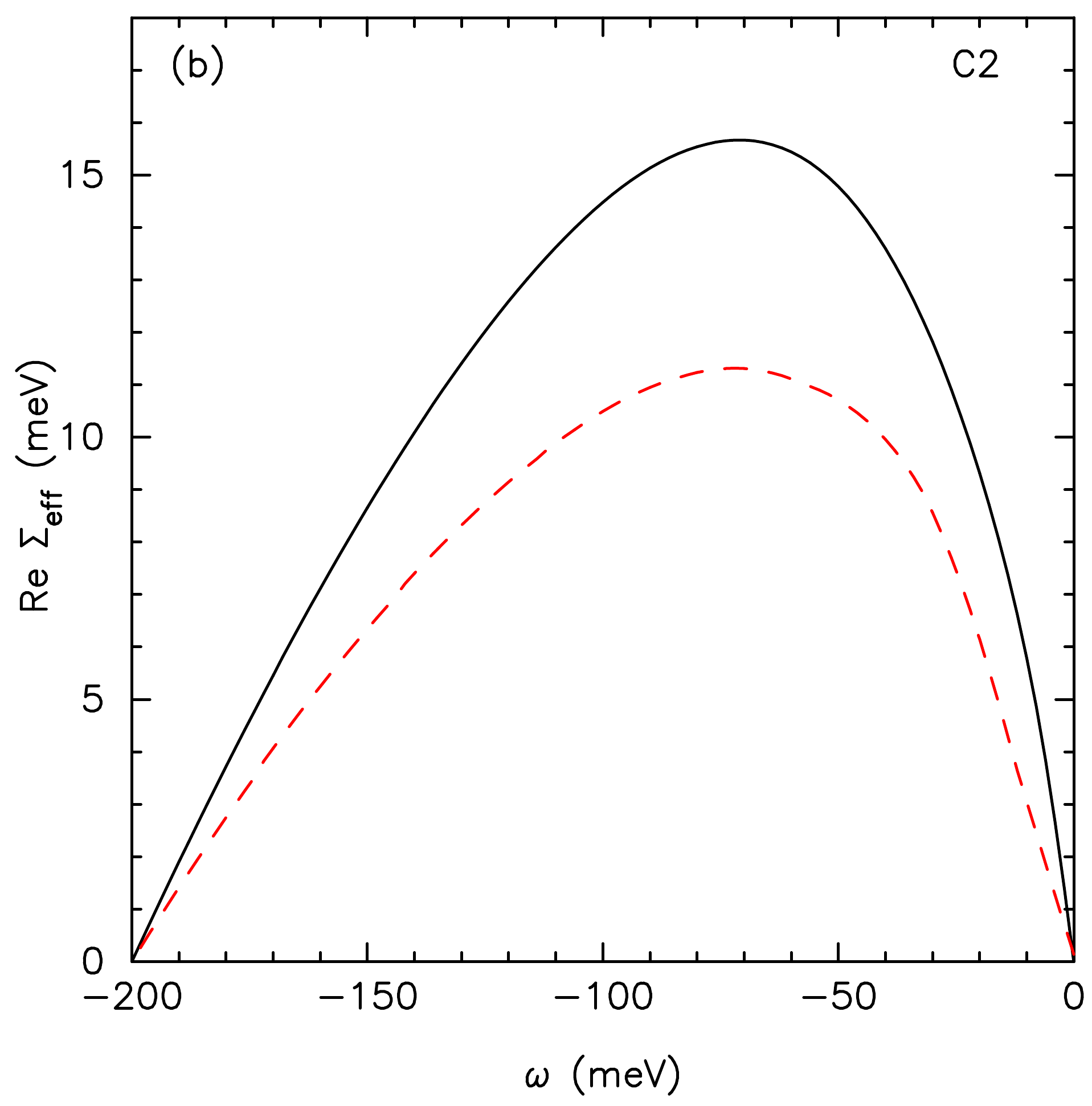}}
\subfigure{
\includegraphics[width=0.45 \columnwidth]{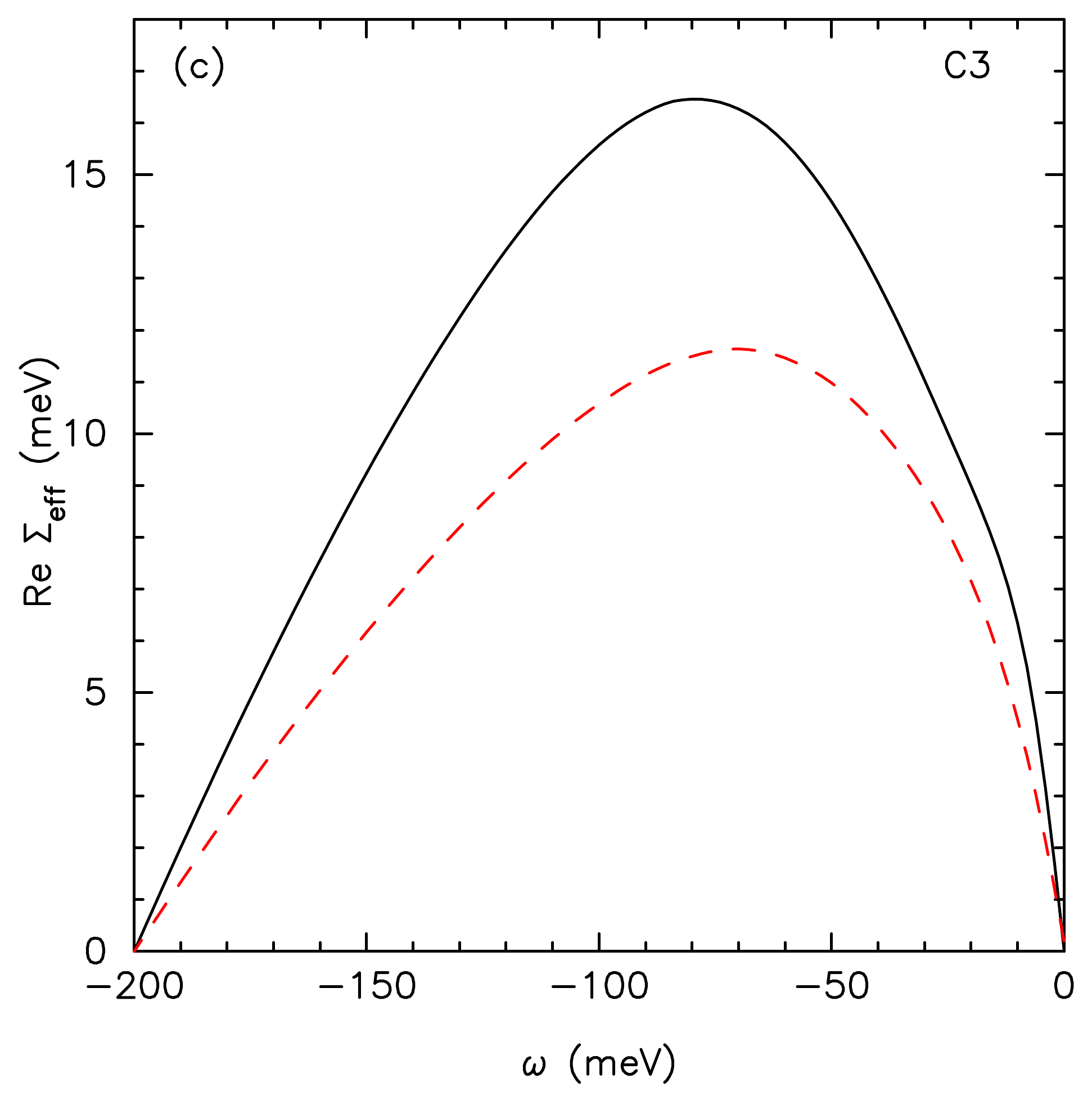}}
\caption{(Color online) Re~$\Sigma_{\rm eff}(k,\omega)$ versus $\omega$ for $x=0.22$ (solid)
and $x=0.30$ (dashed) for (a) the nodal C1 cut, (b) the C2 cut and (c) the
C3 cut.\label{fig:3}}
\end{center}
\end{figure}
The result of this procedure is shown in Fig.~\ref{fig:3}a for the nodal C1 cut
and in \ref{fig:3}b and \ref{fig:3}c for the C2 and C3 cuts. In each case,
the solid curve shows Re~$\Sigma_{\rm eff}(k,\omega)$ for $x=0.22$ and the dashed
curve for $x=0.3$. These results can be compared with Fig.~3b-d by Park \etal\
The fact that Re~$\Sigma_{\rm eff}(k,\omega)$ goes to zero at -200~meV is an 
artifact of the analysis and reflects the choice of the empirical bare band 
which we took to be the same as Park \etal\ 
The effective self energy Re~$\Sigma_{\rm eff}(k,\omega)$
shows a broad peak near $-70$ meV for all three cuts reflecting the kink seen
in the MDC peak of $A(k,\omega)$, Fig.~\ref{fig:2}. The height of this peak is
about half the height of the peak reported in Ref.~\onlinecite{Park}.
Increasing the interaction strength $U/t$ will increase the peak height but 
takes the system into a stronger
coupling regime where the FLEX calculation is less reliable. The C2 and C3
cuts (Figs.~\ref{fig:3}b and c) show a similar behavior. Overall, the results
for Re~$\Sigma_{\rm eff}(k,\omega)$ shown in Fig.~\ref{fig:3} appear similar
to what is seen experimentally. In particular, the peak in Re~$\Sigma_{\rm eff}(k,\omega)$
for $x=0.3$ is reduced by about 30\% from that for $x=0.22$. Thus just as seen
experimentally for the nodal momentum cut and the nearby C2 and C3 cuts, there
is only a modest weakening of the peak in Re~$\Sigma_{\rm eff}(k,\omega)$ when
the doping is increased from 0.22 to 0.3.

\begin{figure}[tbhp]
\begin{center}
\includegraphics[width=0.6 \columnwidth]{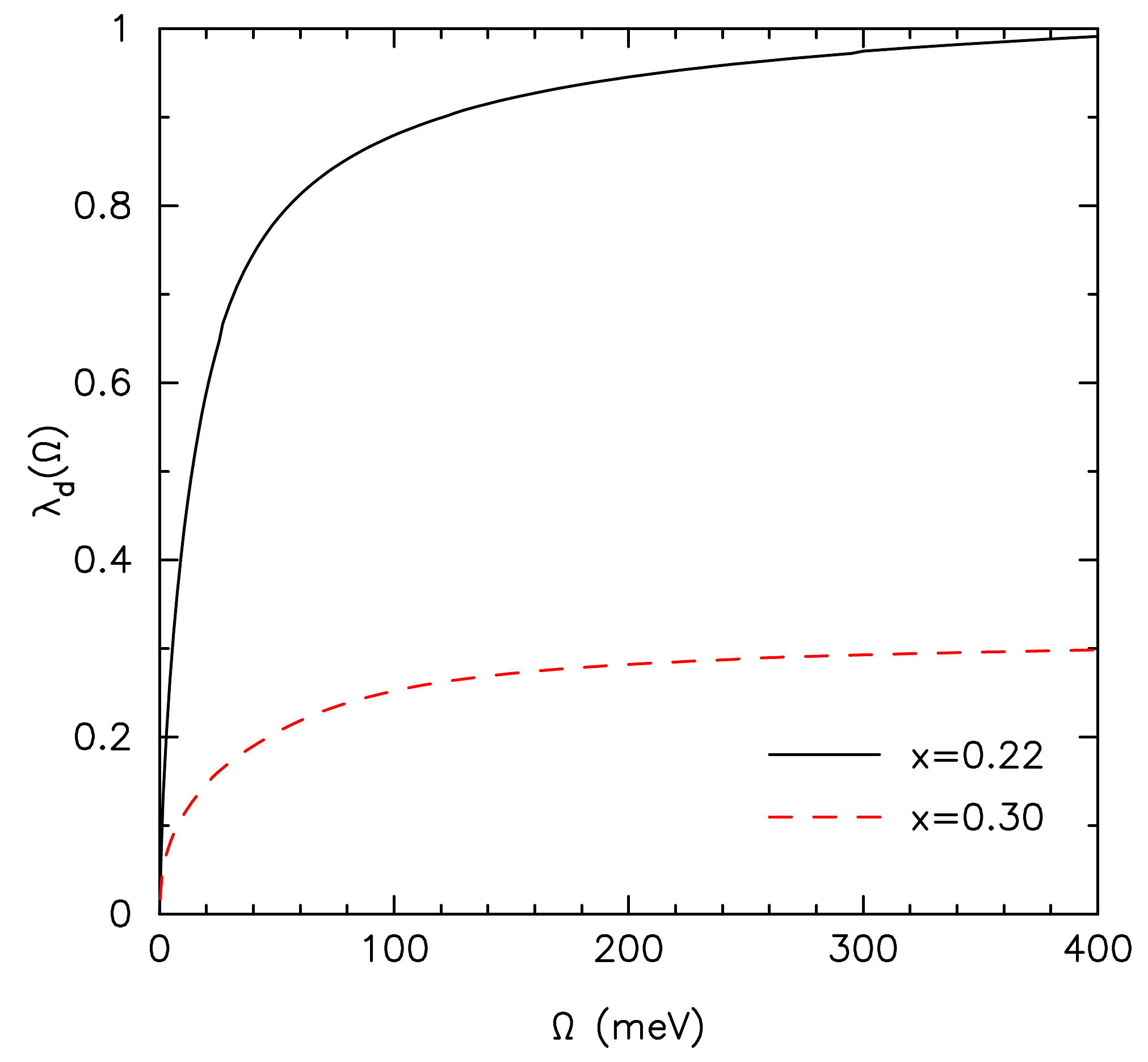}
\caption{(Color online) The $d$-wave pairing strength $\lambda_d(\Omega)$  
obtained from Eq.~(\ref{eq:8}) by cutting off the frequency integration at 
$\Omega$. The solid curve for x=0.22 has a limiting value 1.10 and the 
dashed curve for x=0.3 reaches 0.35.
\label{fig:4}}
\end{center}
\end{figure}

Next we examine the change in the $d$-wave pairing strength when the doping $x$
is increased from 0.22 to 0.3. While the value of $T_c$ depends upon $U$ as
well as the impurity scattering \cite{Kee} associated with the Sr doping,
the change in the $d$-wave pairing strength
\begin{equation}
  \lambda_d=\int_0^{\infty} \frac{d\omega}{\pi\omega}\frac{\left\langle g(k) \;
    \mbox{Im} \; \tilde{\Gamma}(k-k',\omega) \; g(k')
\right\rangle_{k,k'}}{\left\langle g^2(k) \right\rangle_k}
\label{eq:8}
\end{equation}
provides a useful measure of the effect of doping on the pairing. Here
$\tilde{\Gamma}$ is the pairing interaction vertex
\begin{equation}
  \tilde{\Gamma}=\frac{3}{2}\ \frac{U^2\chi_0}{1-U\chi_0}-\frac{1}{2}\ \frac{U^2\chi_0}{1+U\chi_0}
\label{eq:9}
\end{equation}
and $g(k)$ the gap function $(\cos k_x a -\cos k_y a)$. The averages in Eq.~(\ref{eq:8})
are taken over the Fermi surface. Although the charge fluctuation contribution
to the self-energy interaction $\Gamma$ changes sign in the singlet pairing
vertex $\tilde{\Gamma}$, the dominant term in both vertices is the
spin-fluctuation interaction given by the first term in Eqs.~(\ref{Eqflex2}) and
(\ref{eq:9}). 
For $x=0.22$ we find that $\lambda_d=1.10$ while for $x=0.3$ it is
reduced to 0.35. This decrease reflects a redistribution of the spectral weight
from the antiferromagnetic $(\pi,\pi)$ region of the Brillouin zone to other
parts of the Brillouin zone which contribute less or even negatively to the ``$d$-wave"
average in Eq.~(\ref{eq:8}). Thus as the doping $x$ increases, the strength of
the interaction in the anomalous self-energy $d$-wave pairing channel can significantly
decrease while the change in its contribution to the normal self-energy in the
nodal region is relatively modest.
In Fig.~\ref{fig:4} we have plotted $\lambda_d(\Omega)$ obtained
by cutting off the frequency integration in Eq.~(\ref{eq:8}) at $\Omega$. 
This plot shows that the contributions to $\lambda_d$ come from modes with 
energies in a spectral range of several hundred meV. 

{\it Conclusions} - A variety of ARPES studies 
\cite{Zhou, Zhou2, Kordyuk, Yoshida, Park} 
of  hole doped LSCO observe a
50-80~meV kink in the near nodal MDC dispersion which is
reflected as a broad peak in an effective self energy 
Re~$\Sigma_{\rm eff}(k,\omega)$ over an energy range of several
hundred meV. However, there remains disagreement regarding the
origin of this structure and its relationship to the pairing interaction.
Here we have focused on recent results for overdoped LSCO, a system which 
is expected to be more amiable to a weak coupling analysis. In addition, 
this is a single layer cuprate whose Fermi surface at the dopings we have  
studied has been determined by ARPES. 
Within a FLEX approximation we have calculated the
single particle spectral weight for overdoped LSCO and extracted
Re~$\Sigma_{\rm eff}(k,\omega)$ in the same manner as done in the ARPES
experiment. An important ingredient in this calculation was the adjustment
of the bare band parameters such that the renormalized Fermi surface at a 
given doping was fixed to the ARPES determined Fermi surface for that
doping. The results showed that the
renormalized nodal Fermi velocity remained the same and the maximum 
value of Re~$\Sigma_{\rm eff}(k,\omega)$ decreased by $\sim$30\% as 
the doping changed from 0.22 to 0.3, similar to what was seen
experimentally. Using the same parameters, the pairing strength in
the $d$-wave channel was found to decrease from $\lambda_d=1.10$ ($x=0.22$)
to $\lambda_d=0.35$ ($x=0.3$). Thus we conclude that a spin-fluctuation
interaction can give rise to the observed structure in 
Re~$\Sigma_{\rm eff}(k,\omega)$ and also exhibit a significant decrease in the
$d$-wave pairing strength as the doping increases from $x=0.22$ to 0.3.
The point is that the self-energy is determined by the convolution of the
interaction with the spectral weight $A (k ,\omega)$, while the pairing
strength depends on
the $d$-wave projection of the interaction. This $d$-wave projection is more
sensitive to the changes in the momentum dependence of the spin-fluctuation
interaction and the Fermi surface with doping than is the self-energy.
Thus we conclude that the experimental
results reported by Park \etal\ for overdoped LSCO are consistent with a
spin-fluctuation pairing mechanism operating in a spectral range of several hundred meV.

\acknowledgments
The authors thank D.~Dessau for sending them supplementary information for Ref.~\onlinecite{Park}.
DJS acknowledges the support of the Center for Nanophase Materials Science at ORNL,
which is sponsored by the Division of Scientific User Facilities, U.S.~DOE.

\end{document}